\documentstyle[preprint,prc,aps,epsfig]{revtex}
\tighten
\begin{document}
\draft
%%%%%%%%%%%%%%%%%%%%%%%%%%%%%%%%%%%%%%%%%%%%%%%%%%%%%%%%%%%%%%%%%%%%
\title{Meson mass modification in strange hadronic matter}
\author{Subrata Pal, Song Gao, Horst St\"ocker, Walter Greiner}
\address{Institut f\"ur Theoretische Physik, J.W. Goethe-Universit\"at,
D-60054 Frankfurt am Main, Germany}

\maketitle

\begin{abstract}

We investigate in stable strange hadronic matter (SHM) the modification
of the masses of the scalar ($\sigma$, $\sigma^*$) and the vector 
($\omega$, $\phi$) mesons. The baryon ground state is treated in the 
relativistic Hartree approximation in the nonlinear $\sigma$-$\omega$ 
and linear $\sigma^*$-$\phi$ model. In stable SHM, the masses of
all the mesons reveal considerable reduction due to large vacuum 
polarization contribution from the hyperons and small density 
dependent effects caused by larger binding.
\end{abstract}


The study of the properties of hadrons, in particular the light vector
mesons ($\omega$, $\rho$, $\phi$), has recently attracted wide interest
both experimentally and theoretically. Recent experiments from the 
HELIOS-3 \cite{Mas} and the CERES \cite{Wur} collaborations indicate a 
significant amount of strength below the $\rho$-meson peak. This has
been interpreted \cite{Li} as the decrease of the $\rho$-meson mass
in the medium. In an effective chiral model based on the symmetries of 
QCD, Brown and Rho \cite{Bro} predicted an approximate scaling law
for the in-medium decrease of the masses of the mesons and nucleons.
On the other hand, in the quantum hadrodynamic (QHD) model 
\cite{Chi,Ser} based on structureless baryons, the meson masses are
found to increase when particle-hole excitations from the nucleon
Fermi sea are considered. By also including the particle-antiparticle
excitations from the Dirac sea, the meson masses has been, however, 
shown to decrease \cite{Jea,Shi,Sai}.

So far all investigations of in-medium meson mass modification has been
done in the nuclear matter environment. However, presently there is a
growing interest about the possibility of bound strange matter. In analogy
to the stable strange quark matter \cite{Wit}, strange hadronic matter
(SHM) composed of nucleons and hyperons has been speculated \cite{Sch}
to be absolutely stable or at least metastable with respect to weak
hadronic decays.  It is expected that such strange matter may possibly
be created at RHIC and LHC \cite{Spi}. It is therefore worth investigating 
the properties of meson masses in the strange baryonic matter environment.
This study is not only of interest in itself but could serve as a signal
of the formation of SHM in heavy ion collisions. In this letter, in a
complete self-consistent calculation within the framework of the QHD model,
we examine the mass modifications of both, the scalar and the vector 
mesons in SHM.

Let us consider the composition of SHM which is stable with respect to
particle emission. The analysis of level shifts and widths of $\Sigma^-$
atomic level suggest \cite{Dov} a well-depth of $\Sigma$ in nuclear matter
of $U_\Sigma^{(N)} \approx 20-30$ MeV. While for strong processes,
$\Sigma N \to \Lambda N$ and $\Sigma\Lambda \to \Xi N$, the energy released
is $Q \approx 78$ and 52 MeV, respectively. Consequently, systems
involving $\Sigma$'s are unstable with respect to these strong decays.
Analysis of binding energies of $\Lambda$, and emulsion
experiments with $K^-$ beams \cite{Dov} yields well-depths of $\Lambda$
and $\Xi$ in nuclear matter of $U_{\Lambda,\Xi}^{(N)}
\approx 28-30$ MeV. However, in strong decays, $\Lambda\Lambda 
\rightleftharpoons N\Xi$, the system can be stable since $Q \approx 25$
MeV. Therefore, we consider here only the set of baryon species 
$B \equiv \{ N,\Lambda,\Xi \}$, which can constitute stable SHM.

Up to now all investigations of meson mass modification has been performed
in the simple linear Walecka model with nucleons only. Considering here
nonlinear self-interactions of the scalar field $\sigma$ in the QHD model,
the total Lagrangian is
\begin{eqnarray}
{\cal L} &=& \sum_B {\overline \psi}_B \left( i\gamma^\mu\partial_\mu
- m_B + g_{\sigma B}\sigma -  g_{\omega B}\gamma_\mu\omega^\mu \right) 
\psi_B  \nonumber \\
&+& \frac{1}{2}\left(\partial_\mu\sigma\partial^\mu\sigma - m_\sigma^2\sigma^2
\right) + U(\sigma) - \frac{1}{4} \omega_{\mu\nu}\omega^{\mu\nu}
+ \frac{1}{2}m_\omega^2\omega_\mu\omega^\mu + \delta {\cal L} ~,
\end{eqnarray}
where the summation is over all baryon species $B \equiv \{ N,\Lambda,\Xi \}$.
The scalar self-interaction $U(\sigma) = g_2\sigma^3/3 + g_3\sigma^4/4$ 
yields sufficient flexibility to the effective Lagrangian of the model and they
are also necessary for a good reproduction of nuclear ground state properties,
in particular the compressibility. The term $\delta {\cal L}$ contains
renormalization counterterms. This model is not able to reproduce the
observed strongly attractive $\Lambda\Lambda$ interaction. The situation
can be remedied by introducing two additional meson fields, the scalar
meson $f_0(975)$ (denoted as $\sigma^*$ hereafter) and the vector meson
$\phi(1020)$ \cite{Sch} which couple only to the hyperons ($Y$). The
corresponding (linear) Lagrangian is
\begin{eqnarray}
{\cal L}^{YY} &=& \sum_{B=\Lambda,\Xi} {\overline \psi}_B 
\left( g_{\sigma^* B}\sigma^* -  g_{\phi B}\gamma_\mu\phi^\mu \right) 
\psi_B  \nonumber \\
&+& \frac{1}{2}\left(\partial_\mu\sigma^*\partial^\mu\sigma^* 
- m_{\sigma^*}^2\sigma^{*2} \right) - \frac{1}{4} \phi_{\mu\nu}\phi^{\mu\nu}
+ \frac{1}{2}m_\phi^2\phi_\mu\phi^\mu  ~ .
\end{eqnarray}
The self-consistent propagator in the medium for baryon species $B$ can be 
written as sum of the Feynman $[G^F_B(k)]$ and density-dependent $[G^D_B(k)]$ 
parts:
\begin{eqnarray}
G_B(k) &=& G^F_B(k) + G^D_B(k) \nonumber \\
&=& \left(\gamma^\mu k_\mu^* + m^*_B\right) \left[ \frac{1}{k_\mu^{* 2} -
m_B^{* 2} + i\eta} + \frac{i\pi}{E_B^*(k)} \delta\left(k_0^* - E_B^*(k)\right)
\theta\left(k_{F_B} - |{\bf k}| \right) \right] .
\end{eqnarray}
The momentum and energy of baryon $B$ are 
$k_\mu^* = (k_0 - g_{\omega B}\omega_0 - g_{\phi B}\phi_0, {\bf k})$ and 
$E_B^*(k) = [{\bf k}^2 + m_B^{* 2}]^{1/2}$, and $k_{F_B}$ are the Fermi 
momenta. The scalar mean fields shift the mass $m_B$ of the baryons
both in the Fermi and Dirac sea to
\begin{equation}
m_B^* = m_B - g_{\sigma B}\sigma - g_{\sigma^* B}\sigma^* ~.
\end{equation}
In mean field theory (MFT), the total energy of the system
is generated by the presence of all baryons in the occupied Fermi
seas. In contrast, in the relativistic Hartree
approximation (RHA), the effect of the infinite Dirac sea
is also included. The renormalized total energy density in
RHA is given by
\begin{equation}
{\cal E}_{\rm RHA} = {\cal E}_{\rm MFT} + V_B + V_\sigma ~,
\end{equation}
where ${\cal E}_{\rm MFT}$ is the usual MFT energy density \cite{Sch}.
The contribution from vacuum fluctuation of all the baryons is given by
\begin{eqnarray}
V_B = - \sum_B \frac{I_B}{8\pi^2}&\Bigg[& \! m_B^{* 4} \ln \left(\frac{m_B^*}
{m_B}\right) + m_B^3\left(m_B - m_B^*\right) - \frac{7}{2}m_B^2
\left(m_B - m_B^*\right)^2 \nonumber\\
&+& \frac{13}{3}m_B\left(m_B - m_B^*\right)^3
 - \frac{25}{12}\left(m_B - m_B^*\right)^4 \Bigg] ~,
\end{eqnarray}
where $I_B=2I+1$ is the isospin degeneracy of the baryon $B$. The mass shift
of the Dirac sea from $m_B$ to $m_B^*$ produces the vacuum fluctuation
contribution. The renormalized nonlinear $\sigma$-meson 
contribution is \cite{Ser}
\begin{eqnarray}
V_\sigma = \frac{m^4_\sigma}{(8\pi)^2}&\Bigg[& \! 
(1+ \lambda_1+\lambda_2)^2
\ln (1+ \lambda_1+\lambda_2) - (\lambda_1+\lambda_2) \nonumber \\
&-& \frac{3}{2}(\lambda_1+\lambda_2)^2 
- \frac{1}{3}\lambda_1^2(\lambda_1+3\lambda_2)
+ \frac{1}{12}\lambda_1^4 \Bigg] ~,
\end{eqnarray}
where $\lambda_1 = 2g_2\sigma/m^2_\sigma$ and 
$\lambda_2 = 3g_3\sigma^2/m^2_\sigma$.

The meson propagator in the baryonic medium can be computed by summing over
bubbles which consists of repeated insertions of the lowest
order one-loop proper polarization part. This is equivalent to relativistic
random phase approximation (RPA). Since both, scalar ($\sigma$, $\sigma^*$)
and vector ($\omega$, $\phi$) mesons are present in our model, it is 
essential to include scalar-vector mixing which is a pure density dependent
effect. Therefore it is convenient to define a full scalar-vector meson
propagator ${\cal D}_{\rm ab}$ in the form of a $5\times 5$ matrix with indices
$a,b$ ranging from 0 to 4, where 4 corresponds to the scalar meson and 0 to 3
the components of vector meson. Moreover, the (strangeness violating) 
scalar-vector coupling between the strange and nonstrange mesons are 
prohibited by invoking the OZI rule. We therefore consider couplings 
between $\sigma$-$\omega$ and 
$\sigma^*$-$\phi$ separately. We shall present explicitly the calculations only
for the $\sigma$-$\omega$ propagator; the expressions for $\sigma^*$-$\phi$ 
propagator would follow similarly. Dyson's equation for the full 
$\sigma$-$\omega$ propagator ${\cal D}$ can be written in matrix form as
\begin{equation}
{\cal D} = {\cal D}^0 + {\cal D}^0 \Pi {\cal D} ~,
\end{equation}
where ${\cal D}^0$ is the lowest order $\sigma$-$\omega$ meson propagator
\begin{equation}
{\cal D}^0 = \left( \matrix{ D^0_{\mu\nu} & 0 \cr
0 & \Delta^{\!0} } \right) ,
\end{equation}
expressed in terms of the noninteracting $\sigma$ and $\omega$-meson 
propagators
\begin{equation}
\Delta^{\!0} (q) = \frac{1}{q^2_\mu - {\widetilde m}^2_\sigma + i\eta} ~,
\end{equation}
\begin{equation}
D^0_{\mu\nu} (q) = \left( g_{\mu\nu} - q_\mu q_\nu \Big/ m^2_\omega 
\right) D^0(q) ~,
\end{equation}
\begin{equation}
D^0 (q) = \frac{-1}{q^2_\mu - m^2_\omega + i\eta} ~,
\end{equation}
where $q^2_\mu \equiv q^2_0 - {\bf q}$ is the four-momentum carried by the
meson. Note that the effect of nonlinear $\sigma$-interaction (the boson
loops) is to replace the bare meson mass $m^2_\sigma$ in Eq. (10) by 
${\widetilde m}^2_\sigma = m^2_\sigma + \partial^2 U(\sigma)/\partial\sigma^2
= m^2_\sigma + 2g_2\sigma + 3g_3\sigma^2$ (see Ref. \cite{Dob}).

The polarization insertion of Eq. (8) is also a $5\times 5$ matrix
\begin{equation}
\Pi = \left( \matrix{ \sum_B \Pi^B_{\mu\nu}(q) &
\sum_B \Pi^{(M) B}_\nu(q) \cr
\sum_B \Pi^{(M) B}_\mu(q) & \Pi^\pi_\sigma(q) + \sum_B \Pi^B_\sigma(q) } 
\right) ,
\end{equation}
where each entries in Eq. (13) is summed over all the baryons, except
for the pion loop $\Pi^\pi_\sigma (q)$. The effect of relatively large
$\sigma$-width has been accounted by including the contribution of pion
loop to $\Pi_\sigma$ \cite{Sai}; the in-medium modification of the pion loop
is neglected. The pion propagator $\Delta_\pi$ is obtained from Eq. (10)
with ${\widetilde m}^2_\sigma$ replaced by $m^2_\pi$. The pion loop 
polarization contribution to the $\sigma$-meson is
\begin{equation}
\Pi^\pi_\sigma (q) = \frac{3}{2} i g_{\sigma\pi} m^2_\pi
\int \frac{d^4 k}{(2\pi)^4}  \ \Delta_\pi(k) \Delta_\pi(k+q) ~.
\end{equation}
In terms of the baryon propagator the lowest order $\sigma$, $\omega$
and $\sigma$-$\omega$ (mixed) polarizations for the baryon loop $B$ are
respectively given by
\begin{equation}
\Pi^B_\sigma (q) = -i g^2_{\sigma B} \int \frac{d^4 k}{(2\pi)^4} \ 
{\rm Tr} \left[ G_B(k) G_B(k+q)\right] ,
\end{equation}
\begin{equation}
\Pi^B_{\mu\nu} (q) = -i g^2_{\omega B} \int \frac{d^4 k}{(2\pi)^4} 
 \ {\rm Tr} \left[ G_B(k) \gamma_\mu G_B(k+q) \gamma_\nu \right] ,
\end{equation}
\begin{equation}
\Pi^{(M) B}_\mu (q) = i g_{\sigma B} \: g_{\omega B} \int 
\frac{d^4 k}{(2\pi)^4} \ {\rm Tr} \left[ G_B(k) \gamma_\mu G_B(k+q) \right] .
\end{equation}
As in the case for the baryon propagator (Eq. (3)) the above polarization
insertions (except $\Pi^\pi_\sigma$) can be expressed as the sum of
Feynman (F) part and density-dependent (D) part, i.e.
$\Pi^B = {\Pi^B}^{(F)} + {\Pi^B}^{(D)}$. The finite D-part has the form
$G^D\!\times \!G^D + G^D\!\times\! G^F + G^F\!\times \!G^D$ which describes
particle-hole excitations and also includes the Pauli blocking of
$B\bar B$ excitations. Since the polarizations of the D-part
are defined in Ref. \cite{Kur}, we will not refer them here.

The divergent F-part of the polarization insertions can be rendered
finite by adding appropriate counterterms to the Lagrangian of Eq. (1).
For the $\sigma$, each of the baryons loops and the $\pi$-loop have
to be renormalized separately. For any baryon loop contribution to the 
$\sigma$ the usual counterterm Lagrangian \cite{Chi,Ser} is used
\begin{equation}
\delta {\cal L}_\sigma = \sum_{l=2}^4 \frac{\alpha_l}{l\:!} \sigma^l
+ \frac{\zeta_\sigma}{2} \partial_\mu \sigma \partial^\mu \sigma ~.
\end{equation}
The coefficients $\alpha_2$ and $\zeta_\sigma$ can be obtained by imposing
the condition that the propagator in vacuum ($m_B^*=m_B$) reproduces the 
``physical" properties of $\sigma$-meson \cite{Kur}:
\begin{equation}
 {\Pi^B_\sigma}^{\:(RF)}(q^2_\mu; m_B^*=m_B) = \frac{\partial}
{\partial q_\mu^2}{\Pi^B_\sigma}^{\:(RF)}(q^2_\mu; m_B^*=m_B) = 0
~~~~ {\rm at} ~~ q^2_\mu = m^2_\sigma ~.
\end{equation}
The renormalized $\sigma$-meson self-energy for a baryon loop is
\begin{eqnarray}
{\Pi^B_\sigma}^{\:(RF)}(q) &=& \frac{g_{\sigma B}^2 I_B}{2\pi^2}
\Bigg[ \frac{m^2_\sigma - q^2_\mu}{4} + 3m_B\left(m_B^*-m_B\right)
+ \frac{9}{2}\left(m_B^*-m_B\right)^2 \nonumber\\
&-& \frac{3}{2} \int^1_0 dx \ \left(m_B^{* 2} - q^2_\mu x(1-x) \right)
\: \ln \frac{m_B^{* 2} - q^2_\mu x(1-x)}{m_B^2 - m^2_\sigma x(1-x)} \nonumber\\
&-& \frac{3}{2}\left(m_B^{* 2}-m_B^2\right) \int^1_0 dx \ 
\ln \left( 1 - \frac{m^2_\sigma}{m_B^2}x(1-x) \right) \Bigg] ~.
\end{eqnarray}
For the $\pi$-loop we employ the renormalization condition in free
space, ${\Pi^\pi_\sigma}^{\:(RF)}(q^2_\mu)=0$ at 
$q^2_\mu=m^2_\sigma$ \cite{Sai}. We obtain finally
\begin{equation}
{\Pi^\pi_\sigma}^{\:(RF)}(q) = \frac{3 g^2_{\sigma \pi} m^2_\pi}{32\pi^2}
\int^1_0 dx \ \ln \frac{m_\pi^2-q^2_\mu x(1-x)}{m_\pi^2-m^2_\sigma x(1-x)} ~. 
\end{equation}
For the $\omega$, only a wavefunction counterterm ${\cal L}_\omega 
= \zeta_\omega \omega_{\mu\nu} \omega^{\mu\nu}/4$ is required to make
the F-part ${\Pi^B_{\mu\nu}}^{(F)}$ finite. The renormalized $\omega$
self-energy is ${\Pi^B_{\mu\nu}}^{(RF)} = (-g_{\mu\nu}q^2_\mu + q_\mu q_\nu)
{\Pi^B}^{(RF)}$. Employing the renormalization condition in vacuum, 
${\Pi^B}^{(RF)}(q^2_\mu; m_B^*=m_B)=0$ at $q^2_\mu = m^2_\omega$, we obtain
\begin{equation}
{\Pi^B_{\mu\nu}}^{(RF)}(q) = \frac{g^2_{\omega B} I_B}{2\pi^2} q^2_\mu
\int^1_0 dx \ x(1-x) \ln \frac{m_B^{* 2} - q^2_\mu x(1-x)}
{m_B^2 - m^2_\omega x(1-x)} ~.
\end{equation}
As mentioned before, the mixed part $\Pi_\mu^{(M) B}$ of Eq. (17) does
not contribute to vacuum polarization.

The solution of Dyson's equation, Eq. (8), is 
${\cal D} = {\cal D}^0 \big/ (1-{\cal D}^0 \Pi)$. By defining the
dielectric function $\varepsilon$ as
\begin{equation}
\varepsilon = {\rm det} \left( 1-{\cal D}^0 \Pi \right) 
= \varepsilon^2_T \; \varepsilon_L ~ ,
\end{equation}
the poles of the $\sigma$-$\omega$ propagator which define the respective
meson masses in the medium are now contained in the dielectric function
when $\varepsilon=0$. By taking ${\bf q} = (0,0,q)$ where $q=|{\bf q}|$,
we obtained in Eq. (23) the transverse and longitudinal dielectric functions
defined as
\begin{equation}
\varepsilon_T = \left( 1 + D^0 \Pi_T \right) ~,
\end{equation}
\begin{equation}
\varepsilon_L = \left( 1 - \Delta^{\!0} \Pi_\sigma\right) 
\left( 1 - D^0 \Pi_L \right) + \frac{q^2_\mu}{q^2} \Delta^{\!0}D^0 
\left(\Pi_0^{(M)}\right)^2 ~ ,
\end{equation}
where the polarization insertion of Eq. (16) is now split into transverse
$\Pi_T = \Pi_{11} = \Pi_{22}$ and longitudinal $\Pi_L = \Pi_{00} - \Pi_{33}$
components. Because of baryon current conservation, only the 0-th 
component of the mixed part $\Pi^M$ survives. Note that in Eqs. (24)
and (25) the polarizations represent those which have been summed over all
the baryons (see Eq. (13)). The eigencondition for determining the 
collective excitation spectrum (i.e. finding the effective meson masses)
is equivalent to searching for the zeros of the dielectric function. In
particular, for a given three-momentum transfer $q\equiv |{\bf q}|$, the
``invariant mass" of a meson ($\sigma$ or $\omega$) is 
$m^*_m = \sqrt{q_0^2 - q^2}$, where $q_0$ is obtained from the condition
$\varepsilon = 0$. In the present study of meson mass modifications in the
medium, we restrict ourselves to the meson branch in the time-like region
($q^2_\mu>0$).

Since the propagation of the strange $\sigma^*$-$\phi$ mesons are 
decoupled from that of the nonstrange $\sigma$-$\omega$ mesons, we may 
follow the same procedure as given in Eqs. (8)-(25) to obtain the effective
masses of the strange mesons. In particular, a dielectric function (Eq. (23))
is obtained but with the masses of the mesons and their couplings to the 
baryons correspond to the $\sigma^*$ and $\phi$. Moreover, since a linear
$\sigma^*$ interaction is used in this case, $ {\widetilde m}^2_\sigma$ 
in Eq. (10) should be replaced by the bare meson mass $m^2_{\sigma^*}$.

The renormalization conditions in RHA in Eq. (6) is
imposed at $q^2_\mu=0$, while those used in Eq. (19) for $\sigma$ and
$\sigma^*$-mesons are at $q^2_\mu=m^2_\sigma$ and $q^2_\mu=m^2_{\sigma^*}$,
respectively. This difference yields an additional term in the vacuum
fluctuation energy \cite{Kur}, so the total energy density for SHM is 
\begin{equation}
{\cal E} = {\cal E}_{\rm RHA} + \sum_B \frac{a_{\sigma B} + a_{\sigma^* B}}
{4\pi^2} \: I_B \: m_B^2 (m_B^*-m_B)^2 ~,
\end{equation}
where ${\cal E}_{\rm RHA}$ is the RHA energy of Eq. (5) and
\begin{equation}
a_{\sigma B} = \frac{m_\sigma^2}{4m_B^2} + \frac{3}{2} \int_0^1 dx \
\ln \left[ 1 - \frac{m^2_\sigma}{m_B^2} x(1-x) \right] ,
\end{equation}
and a similar expression for $a_{\sigma^* B}$ for $\sigma^*$; for nucleons
$a_{\sigma^* B}=0$.

The field equations are obtained by minimizing the energy density ${\cal E}$
with respect to that field. At a given baryon density $n_B$ and strangeness
fraction $f_S = (n_\Lambda + 2n_\Xi)/n_B$, the set of field equations are  
solved self-consistently in conjunction with the chemical equilibrium 
condition $2\mu_\Lambda = \mu_N + \mu_\Xi$ due to the reaction 
$\Lambda\Lambda \rightleftharpoons N\Xi$. The chemical potential of a baryon
species $B$ is $\mu_B = [k_{F_B} + m_B^{* 2}]^{1/2} + g_{\omega B}\omega_0
+ g_{\phi B}\phi_0$. The four saturation properties of nuclear matter (NM):
density $n_0 = 0.16$ fm$^{-3}$, binding energy $E/B = -16$ MeV, effective
nucleon mass $m_N^*/m_N = 0.78$ and compression modulus $K=300$ MeV are used
to fix the nucleon coupling constants $g_{\sigma N}$,
$g_{\omega N}$ and the parameters $g_2$ and $g_3$ of the 
$\sigma$ self-interaction. The coupling constants for pure NM without hyperons
are shown in Table I. When hyperons are included, i.e. for SHM, they will
contribute to ${\cal E}$ from their vacuum fluctuations $V_B$ even if their 
Fermi states are empty. This entails a redetermination of the coupling constants
for the nucleons. The $V_B$ depends on the effective baryon mass, which
in turn depends on the scalar-baryon coupling constants (see Eqs. (4) and (6)).
Therefore, the $\sigma$ and $\sigma^*$ couplings to the hyperons ($Y$) should 
be predetermined. For this purpose, we adopt the SU(6) model, i.e. 
$g_{\sigma \Lambda}/g_{\sigma N} = 2/3$, 
$g_{\sigma \Xi}/g_{\sigma N} = 1/3$ for the $\sigma$-$Y$ couplings, and
$g_{\sigma^* \Lambda}/g_{\sigma N} = \sqrt{2}/3$, 
$g_{\sigma^* \Xi}/g_{\sigma N} = 2\sqrt{2}/3$ for the $\sigma^*$-$Y$ couplings.
Note that the nucleons do not couple to strange mesons, i.e. 
$g_{\sigma^* N} = g_{\phi N} = 0$. The coupling constants of nucleons for
the SHM are given in Table I. The couplings of $\omega$ to the hyperons 
can be obtained from the well-depth of $Y$ in saturated NM:
$U_Y^{(N)} = g_{\sigma Y}\sigma^{{\rm eq}} - g_{\omega Y}\omega^{{\rm eq}}$
with $U_\Lambda^{(N)} = 30$ MeV and $U_\Xi^{(N)} = 28$ MeV. The $\phi-Y$
couplings are obtained by fitting them to a well-depth
$U_{\Lambda,\Xi}^{(\Xi)} \approx 40$ MeV, for a $\Lambda$ or $\Xi$ in a 
$\Xi$ ``bath" with $n_\Xi \simeq n_0$ \cite{Sch}. To determine the couplings
of the pion to scalar mesons, $g_{\sigma \pi}$ and $g_{\sigma^* \pi}$, we
adjust them to reproduce the widths of $\sigma$ and $\sigma^*$ in free space,
i.e. $\Gamma_s^0 = - \Im {\Pi_s^\pi}^{(RF)}/m_s$ at $q^2_\mu = m_s^2$.
For a conservative estimate we consider $\Gamma_\sigma^0 = 300$ MeV and
$\Gamma_{\sigma^*}^0 = 70$ MeV.

The stability of strange hadronic matter may be explored by considering its
binding energy defined as $E/B = {\cal E}/n_B - \sum_i Y_i m_i$, where
the abundance $Y_i=n_i/n_B$. In Fig. 1, we present the binding energy $E/B$
as a function of baryon density $n_B$ at various strangeness fractions $f_S$.
With increasing $f_S$, the binding energy of SHM is found to increase and the
saturation point is shifted to higher density. This is a consequence of
the opening of new (strangeness) degrees of freedom in the form of hyperons.
At high densities for large $f_S$ 

%%%%%%%%%%%%%%%%%%%%%%%%%%%%%%%%%%%%%%%%%%%%%%%
\vspace{.7cm}

{\small{ TABLE I. The nucleon-meson coupling constants in RHA obtained 
by reproducing
the nuclear matter saturation properties (see text). The results are
for nuclear matter (NM) with strangeness fraction $f_S=0$ and for strange
hadronic matter (SHM) which includes also the vacuum polarization of hyperons.
All the couplings are dimensionless except $g_2$ which has the dimension of
fm$^{-1}$. The bare hadronic masses are $m_N = 939$ MeV, 
$m_\Lambda = 1116$ MeV, $m_\Xi = 1313$ MeV, $m_\sigma = 550$ MeV, 
$m_\omega = 783$ MeV, $m_{\sigma^*} = 975$ MeV, and $m_\phi = 1020$ MeV.}}

\begin{center}
\begin{tabular}{ccccc} \hline\hline

\hfil & \hspace{1cm} $g_{\sigma N} \hspace{1cm} $ 
& \hspace{1cm} $g_{\omega N}$ \hspace{1cm} & \hspace{1cm} $g_2$ \hspace{1cm} 
& \hspace{1cm} $g_3$ \\ \hline
NM& 8.060 & 8.498& 13.962& \hspace{1cm} 1.273 \\
SHM& 7.917& 8.498& 9.471& \hspace{1cm}15.017 \\ \hline\hline

\end{tabular}
\end{center}
%%%%%%%%%%%%%%%%%%%%%%%%%%%%%%%%%%%%%%%%
\noindent values when the Fermi energy of nucleons 
exceeds the effective masses of the hyperons minus their associated 
interaction, the conversion of nucleons at its Fermi surface to hyperons 
of increasingly larger mass is favored. This conversion
lowers the Fermi energy of the nucleons leading to enhanced binding. A stable
SHM with a maximum binding of $E/B = -38.46$ MeV is obtained for a large
$f_S \approx 1.35$ at a relatively high density $n_B \approx 3n_0$. A further
increase of $f_S$ enforces the Fermi energy of the hyperons to increase 
resulting in the decrease of binding. This finding is quite similar to that 
obtained in the quark-meson coupling model \cite{Wan}.

The variation of baryon effective masses $m_B^*$ is shown in Fig. 2 as a 
function of density $n_B$ for varying strangeness $f_S$. It is observed that
$m_B^*$ decreases with increasing $n_B$, with $m_N^*$ having the largest
decrease rate and $m_\Xi^*$ the smallest at each $f_S$ value. Furthermore,
with increasing $f_S$ the effective nucleon mass increases while the
effective masses of hyperons decrease at any density. This effect stems
from decrease of the nonstrange meson fields $\sigma$ and $\omega$ and increase
of the strange meson fields $\sigma^*$ and $\phi$ with increasing $f_S$.
Since the nucleons couple only to $\sigma$ and $\omega$, $m_N^*$ is increased.
The hyperons, however, do couple to all the meson fields resulting in a decrease
of their effective masses with increasing $f_S$, especially for $\Xi$ which
has the strongest coupling to the strange mesons.

In Fig. 3, we display the ``invariant mass" $m_m^*=\sqrt{q_0^2 - q^2}$
of nonstrange $\sigma$-meson, $m_\sigma^*$, (left panels) and $\omega$-meson,
$m_\omega^*$, (right panels) as a function of baryon density $n_B$ for
different strangeness fractions $f_S$. We consider first the features 
observed for $m_\sigma^*$ for small three-momentum transfer 
$q = |{\bf q}| = 1$ MeV (top-left panel). For $f_S=0$, $m_\sigma^*$ is found 
to decrease with increasing $n_B$ for small values
of $n_B \leq n_0$. This reduction is caused due to two competing effects.
The vacuum polarization which leads to reduction of $m_\sigma^*$ dominates
over the density dependent dressing of the meson propagator which causes an
increase in $m_\sigma^*$. In fact, this reduction can be traced back to the
corresponding reduction of $m_N^* < m_N$ in the medium (see Eq. (20)).
However, the decrease in the scalar polarization depends on $m_B^{* 2}$,
$m_B^2$, and $m_\sigma^2$ in a complicated fashion. For densities
above $n_0$, the density-dependent (D) part becomes increasingly dominant
resulting in increase of $m_\sigma^*$. In SHM with $f_S \neq 0$, the
vacuum polarization contribution from the hyperons and the nucleons causes
a considerable suppression of $m_\sigma^*$ at low densities. At higher
densities $m_\sigma^*$ increases, however, the rate of increase above
$\sim n_0$ becomes smaller with increasing $f_S$. An explanation to this
effect is as follows. The D-part of $\sigma$ propagator is primarily
determined by nucleons to which it has the strongest coupling. The increase
of $f_S$ causes an increase in the Fermi momenta of the hyperons while
that of the nucleons decrease. Consequently, with increasing $f_S$ the
decreasing D-part of the nucleons results in a slower rate of increase of
$m_\sigma^*$ with $n_B$. 

At small value of $q=1$ MeV, the (density dependent) effect of scalar-vector
mixing is negligible (see Eq. (25)). On the other hand, for large values of 
$q=500$ MeV (central-left panel) and $q=1$ GeV (bottom-left panel), the mixing 
is more effective.  (For the latter value, the present model may have been 
stretched to the extremes of applicability.) It gives rise to a repulsion 
which again leads to decrease of 
$m_\sigma^*$ at high density. With increasing $f_S$, the onset of this
decrease or the peak positions in the figure are found to shift to higher 
densities. This is a manifestation of the shift of saturation value $E/B$ to 
higher $n_B$ as $f_S$ increases (see Fig. 1).

For small values of $q=1$ MeV, the transverse and longitudinal invariant
$\omega$-meson mass, $m^*_{\omega_T}$ and $m^*_{\omega_L}$ are practically
identical as evident in Fig. 3 (top-right panel). In contrast to $\sigma$ mass,
the vacuum polarization effect is much stronger (see Eq. (22)) and the
density dependent effect is much weaker for the heavier $\omega$ mass. This
causes for $f_S=0$ a substantial decrease of $m_\omega^*$ up to 
$n_B \approx 2n_0$, and a subsequent small increase with density. The reduction 
is more pronounced for SHM where the total vacuum polarization effect is stronger
and the D-part (mainly controlled by N) is weaker. In fact, for $f_S=1.5$,
$m_\omega^*$ decreases considerably even up to large densities. As a 
consequence, the crossing between $m_\sigma^*$ and $m_\omega^*$ at $q=1$
MeV is shifted to lower $n_B$ for the SHM.

At large values of $q=500$ MeV and 1 GeV, the longitudinal mass 
$m^*_{\omega_L}$ (solid line) and transverse mass $m^*_{\omega_L}$ 
(dashed line) get well separated. It is found that $m^*_{\omega_L}$ 
is reduced near nuclear matter density and finally increases with
density attaining values higher than $m^*_{\omega_T}$. As in the
$q=1$ MeV case, the reduction in mass is stronger for SHM. Because of
strong repulsion from the mixing at these high three-momentum values,
$m_\sigma^*$ and $m^*_{\omega_L}$ never cross each other.

In Fig. 4 the ``invariant mass" of the strange $\sigma^*$-meson,
$m^*_{\sigma^*}$, (left panels) and $\phi$-meson, $m_\phi^*$, (right panels)
are shown as a function of baryon density $n_B$ for different strangeness
fractions. In this case the masses are determined by the hyperons as nucleons
do not couple to the strange mesons. As for the nonstrange mesons, the 
masses $m^*_{\sigma^*}$ and $m_\phi^*$ in general decreases with increasing
$n_B$. However, the decrease is more enhanced over the large density range
explored here. This arises due to large vacuum fluctuation contribution
from the hyperons, and, in particular, a small density-dependent part of
the hyperons stemming from the large binding in SHM. At high densities
$n_B \geq 4n_0$, in contrast to nonstrange meson masses, the strange 
meson masses has higher values for larger $f_S$. This reversal in 
behavior for the $m^*_{\sigma^*}$ and $m_\phi^*$ results from the increase
of the Fermi momenta and hence the D-part of the hyperons for large $f_S$.
The scalar mass $m^*_{\sigma^*}$ however becomes sensitive to high values 
of $q$ and drops at $f_S=1.5$ below that for $f_S=0.5$.

We have investigated the meson mass modification in strange hadronic
matter as a function of baryon density for a fixed strangeness fraction.
The results correspond to a metastable SHM. For a given $n_B$, an absolute
stable SHM could be obtained by determining the absolute minimum of the
binding energy $E/B$ as a function of $f_S$. In this situation, we have
found that the masses of the mesons in SHM undergo a drastic reduction
over a wide range of density. Besides the large vacuum part, the minimum
in the energy per baryon $E/B$ at each density enforces a smallest possible 
Fermi momentum and hence the density-dependent for all the baryons.

It is worth mentioning here that in the {\it linear} Walecka model, it has
been demonstrated \cite{Chi,Ser,Jam} that a self-consistent inclusion of 
exchange (Fock) term has an insignificant contribution to binding energy
when the two parameters of this model, $g_{\sigma N}$ and $g_{\omega N}$,
are renormalized to reproduce the same NM saturation properties
of baryon density and binding energy. Even the predicted values of the 
effective nucleon mass, $m^*_N$, and incompressibility, $K$, in this 
Hartree-Fock calculation are almost identical to the Hartree results
\cite{Chi}. However, the applicability of the linear Walecka model to
moderate and high density phenomena can be misleading if the two parameters
$m^*_N$ and $K$  at $n_0$ are not under control. In fact, it was shown
\cite{Gle} that in the {\it nonlinear} Walecka model when the four parameters,
$g_{\sigma N}$, $g_{\omega N}$, $g_2$, and $g_3$ (the latter two are from
the $\sigma$ self-interaction) are fitted to the same NM saturation properties
of $n_0$, $E/B$, $m^*_N$, and $K$ (as in our calculation), nearly all the 
properties obtained in the relativistic Hartree calculation differ from
the mean field results only by $\approx 3\%$ even at $n = 10n_0$. 
(This is in contrast to the linear Walecka model results.) It is
therefore expected that by further inclusion of exchange terms (in all
the baryonic sectors) and pseudoscalar mesons in the present nonlinear 
Walecka model, and performing a self-consistent calculation 
(with the parameters renormalized to the same NM saturation properties), 
the results for $E/B$ and $m^*_B$ will be practically unaltered from the 
Hartree calculation. Consequently, exchange corrections from the nucleonic
and strangeness sectors should also have an insignificant contribution 
and therefore neglected in the present study of the modification of meson 
masses in the medium.

In summary, we have investigated the masses of baryons ($N,\Lambda,\Xi$)
and, in particular, the nonstrange ($\sigma$, $\omega$) and strange
($\sigma^*$, $\phi$) mesons in stable strange hadronic matter. The
ground state properties of the SHM in the relativistic Hartree approximation
is obtained by using a nonlinear $\sigma$-$\omega$ and linear 
$\sigma^*$-$\phi$ Lagrangians. With increasing strangeness fraction, the
effective mass of the nucleons increases while that of the hyperons decreases.
The masses of all the mesons reveal a considerable reduction over a wide
density range with increasing strangeness. This may be attributed to a
large contribution from the vacuum polarization of the hyperons which
causes the decrease in the meson masses at small densities. The larger binding
for the SHM and therefore a smaller density-dependent part helps to 
reduce meson masses at high densities.

\vspace{1cm}

S.P. and S.G. acknowledge support from the Alexander von Humboldt Foundation.

%%%%%%%%%%%%%%%%%%%%%

%%%%%%%%%%%%%%%%%%%%%%%%%%%%%%%%%%%%%%%%%%%%%%%%%%%%%%%%%%
 \newpage 
\vspace{-2cm}

{\centerline{
\epsfxsize=14cm
\epsfysize=17cm
\epsffile{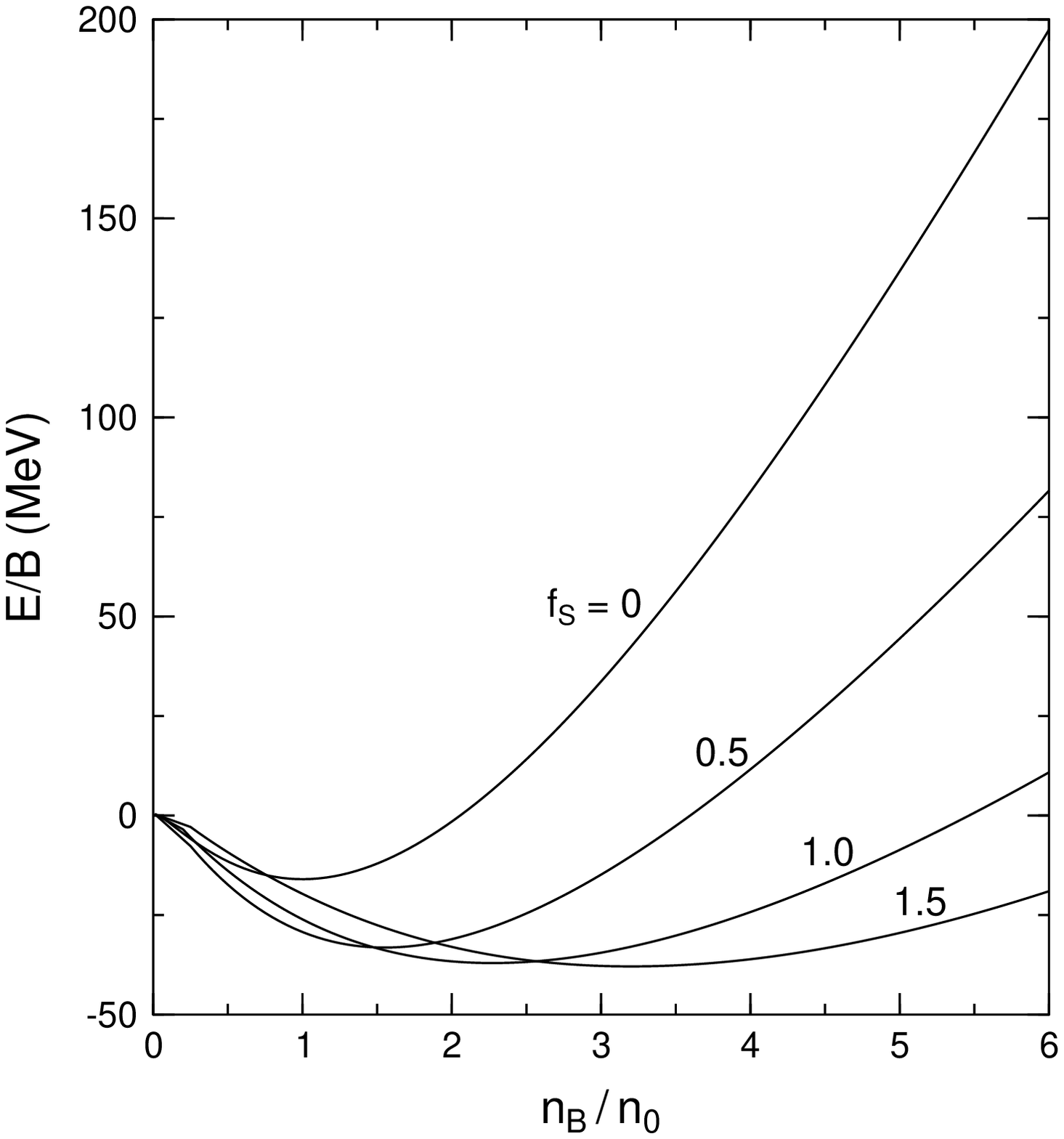}
}}

\vspace{-5cm}

\noindent{\small{
FIG. 1. Binding energy $E/B$ versus baryon density $n_B$ for strange
hadronic matter with various strangeness fraction $f_S$.}}
%%%%%%%%%%%%%
 \newpage 
\vspace{-2cm}

{\centerline{
\epsfxsize=14cm
\epsfysize=22cm
\epsffile{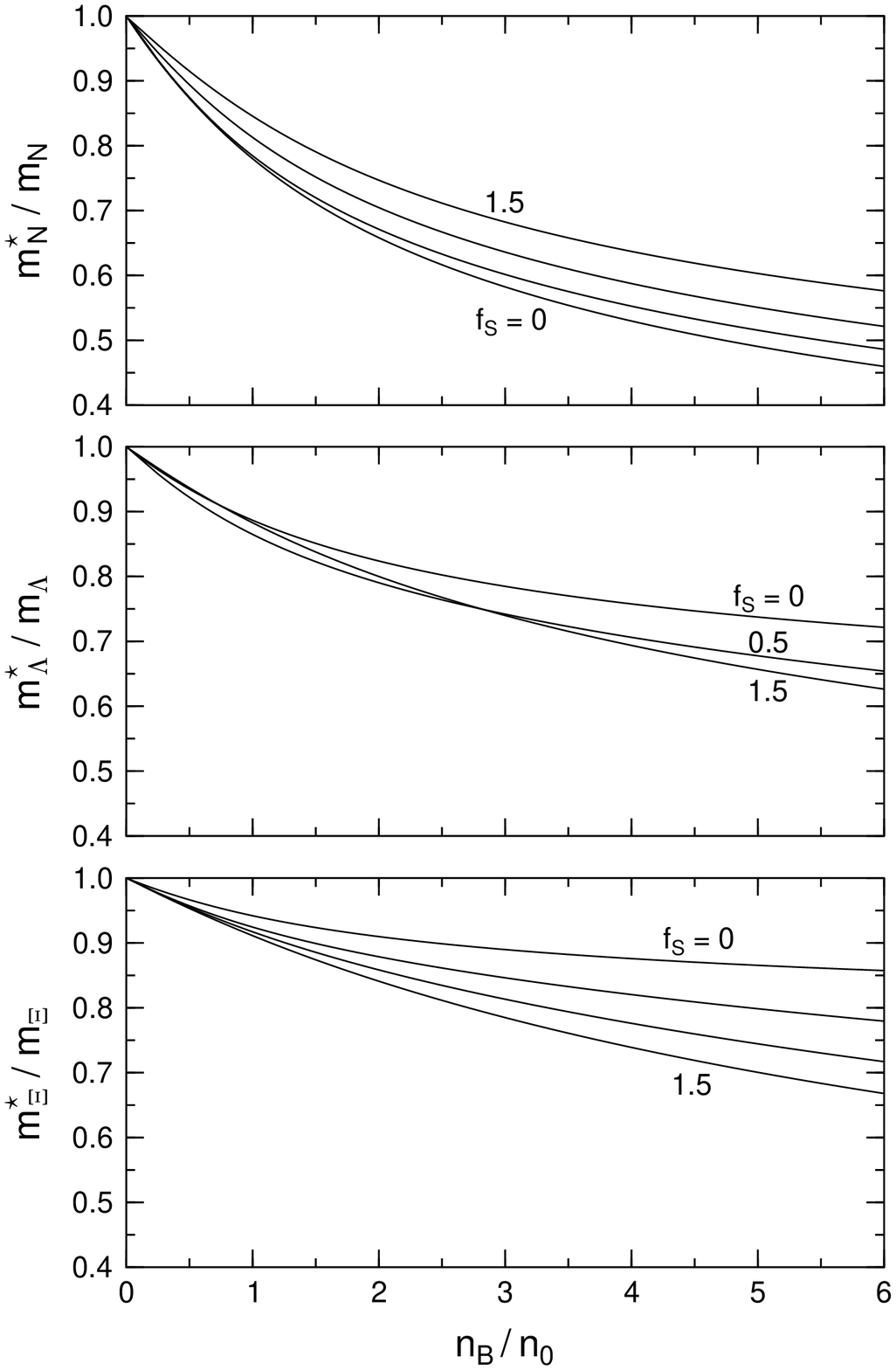}
}}

\vspace{-1.4cm}

\noindent{\small{
FIG. 2. Effective masses of nucleons (top panel), lambda (central panel) 
and cascade (bottom panel) in strange hadronic matter with strangeness fraction
$f_S$ from 0 to 1.5 in steps of 0.5.}}
%%%%%%%%%%%%%%%%
\newpage 
\vspace{-2cm}

{\centerline{
\epsfxsize=14cm
\epsfysize=23cm
\epsffile{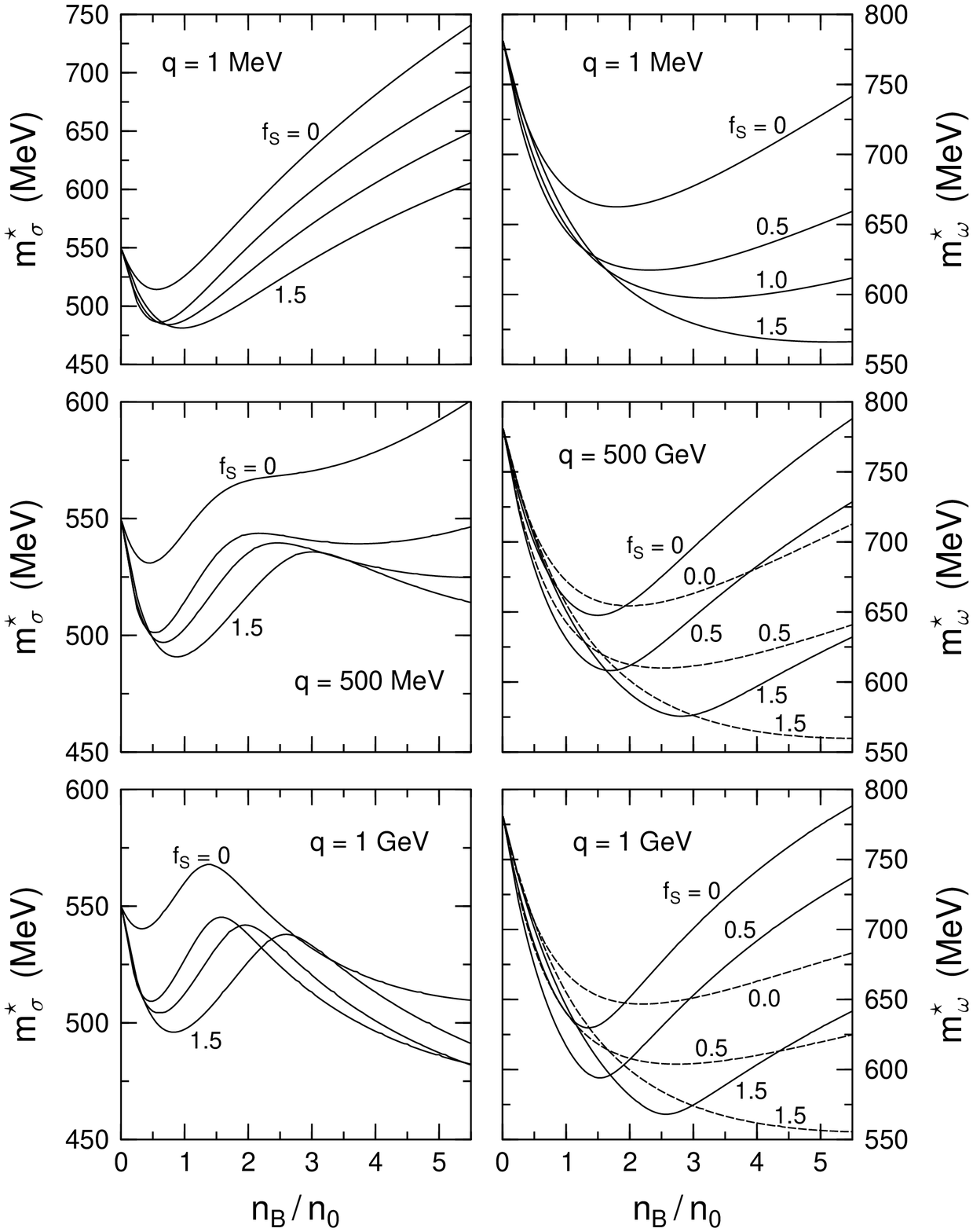}
}}

\vspace{-1.2cm}

\noindent{\small{
FIG. 3. The in-medium masses of $\sigma$-meson (left panels) and $\omega$-meson
(right panels) as a function of baryon density for various $f_S$ with
$q=|{\bf q}| = 1$, 500, and 1000 MeV. The $\sigma$-meson masses are for $f_S$ 
from 0 to 1.5 in steps of 0.5. In the right panels, the solid and dashed lines
correspond to longitudinal and transverse $\omega$-meson masses, respectively.
}}
%%%%%%%%%%%%%%%%
\newpage 
\vspace{-2cm}

{\centerline{
\epsfxsize=14cm
\epsfysize=23cm
\epsffile{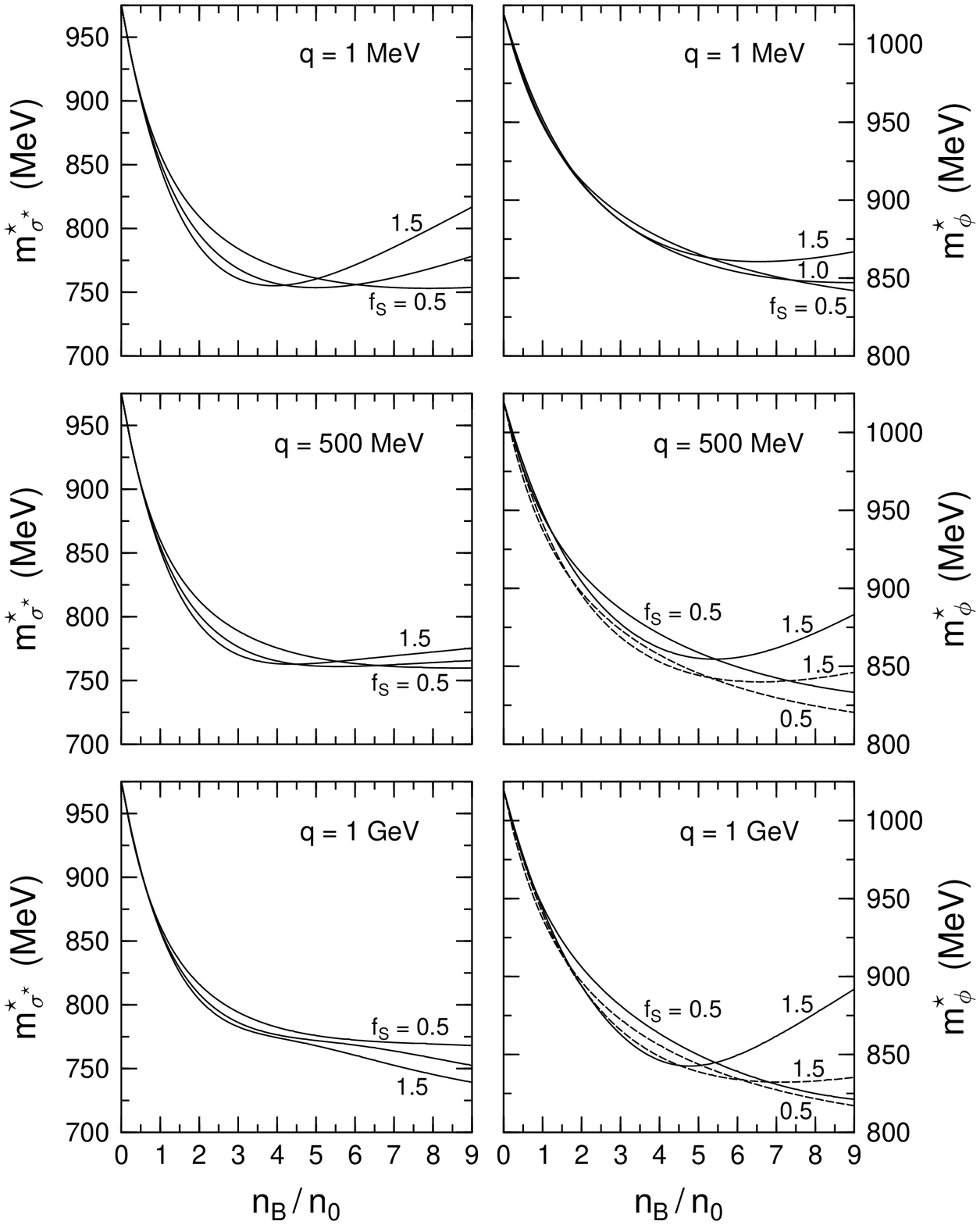}
}}

\vspace{-1.2cm}

\noindent{\small{
FIG. 4. Same as Fig. 3 but for the $\sigma^*$-meson (left panels) and 
$\phi$-meson (right panels).}}

\end{document}